COCCIA LAB

*To discover the causes of social, economic and technological change*

# MEASUREMENT OF ECONOMIC GROWTH, DEVELOPMENT AND UNDER DEVELOPMENT: NEW MODEL AND APPLICATION

Mario COCCIA
ARIZONA STATE UNIVERSITY
Center for Social Dynamics and Complexity
Interdisciplinary Science and Technology Building 1 (ISBT1)
550 E. Orange Street, Tempe- AZ 85287-4804 USA

and

CNR -- NATIONAL RESEARCH COUNCIL OF ITALY
Via Real Collegio, 30-10024, Moncalieri (TO), Italy

*E*-mail: mario.coccia@cnr.it

# MEASUREMENT OF ECONOMIC GROWTH, DEVELOPMENT AND UNDER DEVELOPMENT: NEW MODEL AND APPLICATION

MARIO COCCIA[1]

ARIZONA STATE UNIVERSITY &
CNR -- NATIONAL RESEARCH COUNCIL OF ITALY

*E*-mail: mario.coccia@cnr.it

**Abstract**. This paper presents a simple model to measure the relative economic growth of economic systems. The model considers *S*-Shaped patterns of economic growth that, represented with a linear model, measure how an economic system grows in comparison with another one. In particular, this model introduces an approach which indicates if the economic system has a process of economic growth, development or under development. The application of the model is provided for regions and macro regions of the Italian economic system.

[1] **Acknowledgements**. I am grateful to Michele Mininni (University of Bari, Italy), Luigi Montrucchio (University of Torino, Italy), Alessandro Flamini (Graduate Institute of International Studies-Geneva, Switzerland), David Audretsch (Max Planck Institutes, Jena Germany), Jagannadha Pawan Tamvada (Max Planck Institutes, Jena Germany), and Angelo Reati (European Commission, Brussels) for valuable suggestions and discussion. I add that the responsibility for all views expressed is entirely mine.

## MODEL

Let *Y(t)* be the total output at time *t* of the economic system *Y'* and *X(t)* be the total output at the same time of the economic system *X'*;

let Y'⊇X', $b_1$ and $b_2$ be the rates of growth of total outputs *Y* and *X*, respectively, such that $B_1 = \frac{b_2}{b_1}$;

if *Y* and *X* increase, in the long run, according to some *S*-shaped pattern of growth, then $B_1 = \frac{b_2}{b_1}$ measures the relative economic growth of the economic system *X'* in relation to the economic growth Y'.

In fact, if both *Y* and *X* increase in the long run according to some *S*-shaped pattern of growth (Lewis, 1955; Jarne *et al.*, 2005), one way to represent such a pattern formally is in terms of the differential equation of the well-known logistic function. In the case of *Y(t)* we have:

$$\frac{1}{Y}\frac{dY}{dt} = \frac{b_1}{K_1}(K_1 - Y)$$

This equation can be rewritten as

$$\frac{K_1}{Y}\frac{1}{(K_1 - Y)}dY = b_1 dt \qquad \frac{dY}{Y} - \frac{-dY}{(K_1 - Y)} = b_1 dt$$

Upon integrating we obtain

$$\log Y - \log(K_1 - Y) = A_1 + b_1 t$$

$$\log\frac{K_1 - Y}{Y} = a_1 - b_1 dt$$

$$Y = \frac{K_1}{1 + \exp(a_1 - b_1 t)}$$

where $a_1 = b_1 dt$, and $t_1$ is the abscissa of the point of inflection.

Thus the growth of Y and X can be described as:

$$\log\frac{K_1 - Y}{Y} = a_1 - b_1 t \qquad [1]$$

for *X(t)* we proceed in similar way of *Y(t)* and we have:

$$\log\frac{K_2 - X}{X} = a_2 - b_2 t \qquad [2]$$

respectively.

It can be readily verified that the logistic curve is a symmetrical *S*-shaped curve with a point of inflection at 0.5K[2]

Solving the equations [1] and [2] for *t*,

$$t = \frac{a_1}{b_1} - \frac{1}{b_1}\log\frac{K_1 - Y}{Y} = \frac{a_2}{b_2} - \frac{1}{b_2}\log\frac{K_2 - X}{X}$$

which immediately yields the expression

$$\frac{Y}{K_1 - Y} = C_1\left(\frac{X}{K_2 - X}\right)^{\frac{b_1}{b_2}} \qquad [3]$$

Clearly:

$C_1 = \exp\left(\frac{a_2 b_1 - a_1 b_2}{b_2}\right)$, which can be written in a simplified form as

$C_1 = \exp[b_1(t_2 - t_1)]$ since, as noted earlier, $a_1 = b_1 t_1$ and $a_2 = b_2 t_2$ (cf. Eqs. [1] and [2]).

When X and Y are small in comparison with their final value, Eq. [3] reduces to

$$\frac{Y}{K_1} = C_1\left(\frac{X}{K_2}\right)^{\frac{b_1}{b_2}}$$

Hence the following simple model of economic growth is obtained $X = A_1(Y)^{B_1}$ [4]

where $A_1 = \frac{K_2}{(K_1)^{\frac{b_2}{b_1}}} C_1$ and $B_1 = \frac{b_2}{b_1}$

The Eq. [4] was used by Huxley (1932) to describe the shape changes which animals and plans undergo during growth. In similar way this allometry equation [4] can be used to describe the changes of economic systems undergo during economic growth. The standard approach is to submit relative growth data to a logarithmic transformation before carrying out calculations. In fact, the logarithmic form of the equation $X = A_1 (Y)^{B_1}$ is a simple linear relationship,

$$\log X = \log A_1 + B_1 \log Y$$

[2] Briefly, $a_1$ is a constant depending on the initial conditions, $K_1$ is the equilibrium level of growth, and $b_1$ is the rate-of-growth parameter.

$B_1$ is the allometry exponent of the $X$ relatively to the $Y$.

If the relative growth of the two dimensions were *isometric* (i.e. it produces economic growth), the allometry exponent $B_1$ should have a unit value. This hypothesis is expressed as:

$$B_1 = 1$$

On the other hand, the hypothesis that X increases at greater relative rate that Y, the hypothesis of *positive allometric growth* or *economic development*, could be expressed as:

$$B_1 > 1$$

the hypothesis that X a negative allometric growth (*under development*) relatively to Y would be expressed as:

$$B_1 < 1$$ ■

*Remark:* Gompertz function

It is not necessary that the growth curves of economic system be of the logistic form. For instance, suppose that the pattern of economic growth is described as a Gompertz function:

$$\frac{1}{Y}\frac{dY}{dt} = J_1 \exp(I_1 - J_1 t)$$

The integral form of this equation is:

$$\log Y = -\exp(I_1)\exp(-J_1 t) + \log K_1$$

or $Y = K_1 \exp[-\exp(I_1 - J_1 t)]$

Thus the growth of Y and X can be described as:

$$\log\log\frac{K_1}{Y} = I_1 - J_1 t \qquad [5]$$

and

$$\log\log\frac{K_2}{X}=I_2-J_2t \qquad [6]$$

This equation represents the Gompertz type of economic growth process and unlike the logistic, the Gompertz curve[3] is an asymmetrical S-shaped curve with a point of inflection at $Y=K/e$ of the limit to growth.

Solving the equations [5] and [6] for *t*, we have

$$t=-\frac{1}{J_1}\left(\log\log\frac{K_1}{Y}-I_1\right)=-\frac{1}{J_2}\left(\log\log\frac{K_2}{X}-I_2\right)$$

or

$$\log\log\frac{K_1}{Y}=\log M+\log\left(\log\frac{K_2}{X}\right)^v \qquad [7]$$

where:

$v=\frac{J_1}{J_2}$, $M=\frac{G_1}{G_2^v}$, $G_1=e^{I_1}$, $G_2=e^{I_2}$; solving for *Y*, we have

$$Y=K_1\exp\left[-M\left(\log\frac{K_2}{X}\right)^v\right]$$

If both variables under consideration grow at the same rate, that is, *v=1*, then

$$Y=\left(\frac{K_1}{K_2{}^M}\right)X^M \qquad [8]$$

Further, $M=\frac{e^{I_1}}{e^{I_2}}$, since *v=1*. Thus the value of *M* depends only on two constants; it does not involve the rates of growth of the variables under consideration. Eq. [8] is, of course, identical with the earlier model given by Eq. [4] since it can be rewritten as

$$X=A_2(Y)^{B_2} \qquad [9]$$

where $A_2=\frac{K_2}{K_1^{\frac{1}{M}}}$, $B_2=\frac{1}{M}$.

[3] Briefly, $I_1$ is a constant, $K_1$ is the equilibrium level of growth, and $J_1$ is the rate-of-growth parameter.

*Remark:* generalization

Suppose that X and Y increase according to different forms of *S*-shaped curves, let

$$\frac{1}{Y}\frac{dY}{dt}=u_1\left(K_1-Y\right) \qquad \frac{1}{X}\frac{dX}{dt}=u_2\left(K_2-X^m\right)$$

where $u_1=\frac{b_1}{K_1}$ and $u_2=\frac{b_2}{K_2}$. The solution of this system of equations is given by

$$\frac{1}{X}\frac{dX}{dt}=B_3\left(\frac{1}{Y}\frac{dY}{dt}\right)$$

or $X=A_3(Y)^{B_3}$ [10]

where $\log A_3$ is a constant of integration and

$$B_3=\frac{1}{m} \qquad m=\frac{K_1u_1}{K_2u_2} \qquad A_3=\left(\frac{K_2}{K_1}\right)^{\frac{1}{m}}.$$

Clearly, the form of the Eq. [10] is identical with that of Eq. [4] and [9].

## APPLICATION

### Measuring the morphological change of economic growth

The application of the model above is based on Italy. This study uses the variable: region's *i* annualized grow rate of per capita Gross Domestic Product (GDP), measured in Euros (value in 2003; cf., table 1A). The source data are from ISTAT (Italian National Institute for Statistics) for the 1980-2003 period, divided into 20 regions and three macro regions (Table 1).

Table 1 - Italian regions and macro regions

| *Abbreviations* | *Macro-Region* | *Regions* |
|---|---|---|
| N Italy | North of Italy | Liguria, Lombardia, Piemonte and Valle d'Aosta; Emilia-Romagna, Friuli-Venezia Giulia, Trentino-Alto Adige, Veneto |
| C Italy | Central part of Italy | Abruzzo, Lazio, Marche, Toscana, Umbria |
| S Italy | South of Italy and Islands | Basilicata, Calabria, Campania, Molise, Puglia, Sardegna, and Sicilia |

The study here measures and analyzes the patterns of economic growth of some Italian regions (which are not included in the North of Italy), in relation to (rich) Northern Italy. Moreover, the paper investigates the economic growth of Italian macro-regions, Centre and South of Italy, in relation to the North of Italy.

Although differences in technology, preferences, and institutions do exist across regions, these differences are likely to be smaller than those across countries. Firms and households of different regions within a single country tend to have access to similar technologies and have roughly similar tastes and cultures. Furthermore, the regions share a common central government and therefore have similar institutional setups and legal systems. This relative homogeneity means that absolute convergence is more likely to occur across regions within countries than across countries. Another consideration is that inputs tend to be more mobile across regions than across countries. Legal, cultural, linguistic, and institutional barriers to factor movements tend to be smaller across regions within a country than across countries.

The following assumptions are the basis of the model:

1. let $S_1$= regional system 1; $S_2$= regional system 2, MR=Macro region*, I assume that $S_{1,}$ $S_{2,}$ MR $\subseteq$ A (country system), $S_1 \cap MR = \varnothing$ and $S_2 \cap MR = \varnothing$
2. patterns of per capita GDP grow in the short-medium run with an *S*-Shaped pattern (Jarne *et al.*, 2005; *see* figure 1A).
3. adjacent regions within a country (e.g., Italy) are homogenous in terms of groups of people, institutions and firms, investment habits, savings, consumption, social status, cultures, tastes, financial positions, open-mindedness, laws, industries, etc.
4. the North of Italy is a richer economic system than other Italian regions and macro-regions and it is able to promote economic growth in other regions.

*Remark*: Although the North of Italy is formed by the North East and North West, which have different characteristics, I assume Northern Italy to be a homogeneous system (on the whole) richer than other Italian regions.

*Remark:* This analysis performs comparison among homogenous elements which are sub-sets of a system; in other words, given the $C_1$ and $C_2$ country systems and the $R_1$ region such that $R_1 \subseteq C_1$, we cannot measure the economic growth of $R_1$ in relation to $C_2$ because $R_1 \not\subset C_2$. (i.e., the economic growth of Italian regions cannot be compared with a German region because of differences in habits, institutions, laws, national system of innovation, etc.; therefore, we draw a comparison "within" the country).

The analysis is carried out on the matrix R= 24 (years) × 13 (regions: 12 regions+1 macro-region* of North Italy) of the Italian economic system:

$$R = \begin{bmatrix} GDP\ 1980region_1 & .... & GDP\ 1980region_i & ... & GDP\ 1980region_{12} & GDP\ 1980Macro-region^* \\ ... \\ GDP\ year\, i/region_1 & .... & GDP\ year\ i/region_i & ... & GDP\ year\ i/region_{12} & GDP\ year\ i/Macro-region^* \\ ... \\ GDP2003region_1 & .... & GDP\ 2003region_i & ... & GDP\ 2003region_{12} & GDP\ 2003Macro-region^* \end{bmatrix}$$

and the matrix MR= 24 (years) ×3 (Macro Regions):

$$MR = \begin{bmatrix} GDP\ 1980macro-region_1 & GDP\ 1980macro-region_2 & GDP1980macro-region_3^* \\ ... \\ GDP\ year\ i/macro-region_1 & GDP\ year\ i/macro-region_2 & GDP\ year\ i/macro-region_3^* \\ ... \\ GDP\ 2003macro-region_1 & GDP\ 2003macro-region_2 & GDP\ 2003macro-region_3^* \end{bmatrix}$$

*Remark:* Macro region* is the North of Italy, which I assume to be the engine of overall Italian economic growth; *region*$_i$ and other macro-regions are not included in the Macro region*, *i.e.* the North of Italy.

Our function of economic growth is given by:

$$y_t = a \cdot (x_t)^B$$

where:

$a$ is a constant

$y_t$ is region *i*'s annualized growth rate of GDP per capita at time *t*

$x_t$ is Northern Italy's annualized growth rate of GDP per capita at time *t*, such that $t \in \{1980, ..., 2003\}$, measured in Euros (2003 value).

The North of Italy is a driving force (similar to a locomotive) of the economic growth of other Italian regions (wagons) and of the overall Italian economic system.

*Remark*: I assume, as already said, that growth in a leading region is capable of promoting growth in other regions to a lower, similar, or higher extent.

*Remark*: Let $b_1$ and $b_2$ be the growth rates of the total outputs of economic systems, *Y (region)* and *X (North of Italy)* respectively, so that

$B_1 = \frac{b_1}{b_2}$ measures the relative economic growth of economic system *Y* in relation to the economic growth of *X*.

The Moving Average is applied to data to smooth out the fluctuations underlying the GDP growth rate in order to eliminate the serial correlations. In particular, the moving averages are calculated over three years, instead of five years, to avoid having shorter time series for the regression analyses. These data are then transformed into natural logarithmic data before calculations are carried out to have normal distribution.

In fact, the logarithmic form of equation $y_t = a \cdot x_t^{B}$ is a simple linear relationship:

$$\ln y_t = \ln a + B \ln x_t + \varepsilon_t \qquad [11]$$

I consider the following hypotheses:

- if the GDP per capita growth of the region represented on the *y*-axis is lower than the GDP per capita growth of the macro region* represented on the *x*-axis and B = *1*, both total outputs (regions and North of Italy) are growing at the same rate (*economic isometric growth of regional systems*); in this case regional economic growth follows steady-state growth (s*). This is similar to a balanced-growth path.

- if the GDP per capita growth of the region represented on the *y*-axis is lower than the GDP per capita growth of the macro region* represented on the *x*-axis and B < *1*, the component represented on the *y* axis (growth rate of regional GDP per capita) is growing more slowly than the component on the *x* axis (which represents the North of Italy); this hypothesis (Hp) presents a negative disproportionate (allometric) growth of the regions in relation to the Macro region* (or under development). In other words, this generates a particular kind of unbalanced growth.

- if the GDP per capita growth of the region represented on the *y*-axis is lower than the GDP per capita growth of the Macro region* represented on the *x*-axis and $B > 1$, the *y* axis component is growing faster than the *x* axis component; there is a positive disproportionate (allometric) growth or *economic development* of the regional systems in relation to the Macro region*.

*Remark*. The concepts of isometric, negative, and positive disproportionate (allometric) economic growth are based on an unconditional approach, which does not consider additional variables.

*Remark*. If there are economic systems $\Omega$ and $\Theta$ (e.g. regions), with $\Omega, \Theta \subseteq A$ (country system: e.g. Italy), represented on the *y*-axis, where $\Omega$ is a poor region and $\Theta$ a rich one (i.e., with high GDP per capita), and $X \subseteq A$, an economic system richer than $\Omega$ and $\Theta$; if *X* is represented on the *x*-axis, the patterns of economic growth of $\Omega$ towards $\Theta$ depend on parameter B of $\Omega$ and $\Theta$ respectively.

*Remark*. If there are several economic systems $\Omega_1, \Omega_2, \ldots, \Omega_i, \ldots, \Omega_n \subseteq A$ represented on the *y*-axis, with different starting situations represented by GDP per capita, and $X \subseteq A$, a rich economic system, if *X* (represented on *x*-axis) is the engine of the economic growth of the overall system A, this model shows the patterns of economic growth of these economic systems $\Omega_i$ ($i=1, \ldots n$) in relation to each other regions and to the rich economic region *X*.

To sum up, the cases analyzed in this study are the Italian regions and two macro-regions over 24-year time series. As the patterns of economic growth of Italian regions and macro regions in the period 1980-2003 are *S*-shaped functions (*see* figure 1A in the Appendix), I apply model [11].

The model [11] has linear parameters that are estimated by the Ordinary Least-Squares Method. The calculation is carried out using the SPSS statistics software. Table 2 presents the typology of economic growth of regions and macro-regions in relation to the North of Italy.

Table 2 - Typology of economic growth of Italian regions and macro-regions (period 1980-2003)

| *Regions* | $\hat{B}_1$ *(Std. Err.)* | *Typology of economic growth with sig. 5%* | $R^2$ *adj* |
|---|---|---|---|
| Sardegna | 0.582* (0.230) | ALLOMETRY – | 0.213 |
| Marche | 0.697** (0.102) | ALLOMETRY – | 0.684 |
| Campania | 0.798# (0.340) | suspect Allometry – | 0.184 |
| Basilicata | 0.294** (0.224) | ALLOMETRY – | 0.036 |
| Puglia | 1.027# (0.156) | suspect Allometry + | 0.689 |
| Molise | 0.793* (0.069) | ALLOMETRY – | 0.855 |
| Lazio | 0.755# (0.170) | suspect isometry | 0.472 |
| Calabria | 0.357** (0.108) | ALLOMETRY – | 0.321 |
| Abruzzo | 0.628** (0.110) | ALLOMETRY – | 0.622 |
| Umbria | 0.618# (0.122) | suspect isometry | 0.533 |
| Toscana | 0.651** (0.083) | ALLOMETRY – | 0.744 |
| Sicilia | 0.261** (0.100) | ALLOMETRY – | 0.234 |
| **Macro-Regions** | | | |
| *C Italy* | *0.630*** (0.056) | ALLOMETRY – | 0.857 |
| *S Italy* | *0.927#* (0.105) | suspect isometry | 0.778 |

*Note:* *Level of significance 5% ** Level of significance 1‰ # A non-significant B means that there is no (significant) relationship between the growth rate of the region under investigation and Northern Italy. Statistical inference on regression coefficients indicates suspect isometry or negative/positive disproportionate (allometric) growth: the Null Hypothesis is $H_0 : \hat{B} = 1$

The parametric estimates are presented in Tables 2A-3A (in the Appendix).[4] Firstly, in a few cases the results from the Durbin-Watson (DW) test indicate the presence of serial correlation in the residuals of the equations. In the other cases, however, the DW's *d* statistic is in the zone of indeterminacy and of acceptance of the null hypothesis. The parametric estimates of the models are unbiased, the *t*-test presents significance of the coefficients at 1‰ and 5%, and the $R^2$ values are high, except for a few cases. Thus, in the majority of cases the models explain more than 50% of variance in the data. The results reveal (with 5% level of significance) that eight regions have negative disproportionate (allometric) economic growth; moreover, there is one suspect negative allometry (Campania), two suspect isometric growth (Lazio and Umbria) and one suspect positive allometry or economic development (Puglia). These results are due to a non-significant B in the regression analysis, even if the *t*-test B=1 provides for acceptance or rejection of the null hypothesis.

The relative changes in the economic growth of two macro-regions in relation to the growth rate of the North of Italy suggest that there is a (suspect) isometric economic growth of the South of Italy (coefficient 0.93) and a negative economic allometry for the central part of Italy in relation to the North of Italy over the 1980-2003 period. The functions of economic growth of these two Italian macro-regions are:

$$CITALYy_t = 0.603 \cdot NITALY\ x_t^{0.63}$$

(C Italy with *negative* disproportionate – allometric –economic growth)

$$SITALYy_t = 1.135 \cdot NITALY\ x_t^{0.93}$$

(S Italy with suspect "isometric" economic growth)

[4] To reduce the autocorrelation of Calabria, Lazio, Toscana and the central part of Italy, these areas are standardized to the 1981 value, while Sardegna and Umbria to the 2003 value.

Therefore, this model suggests three behaviors (Figure 1):

- *Economic isometric growth* (suspect) of the regional system if both growth rates of GDP per capita (regions and North of Italy) are growing at the same rate.
- *Economic development of the regional system* if the regional growth rate of GDP per capita is growing faster than the growth rate of GDP per capita in the North of Italy.
- *Negative economic disproportionate (allometric) growth of the regional system* if the growth rate of the GDP per capita is growing more slowly than the growth rate of GDP per capita in the North of Italy.

Figure 1 – Patterns of spatial economic growth

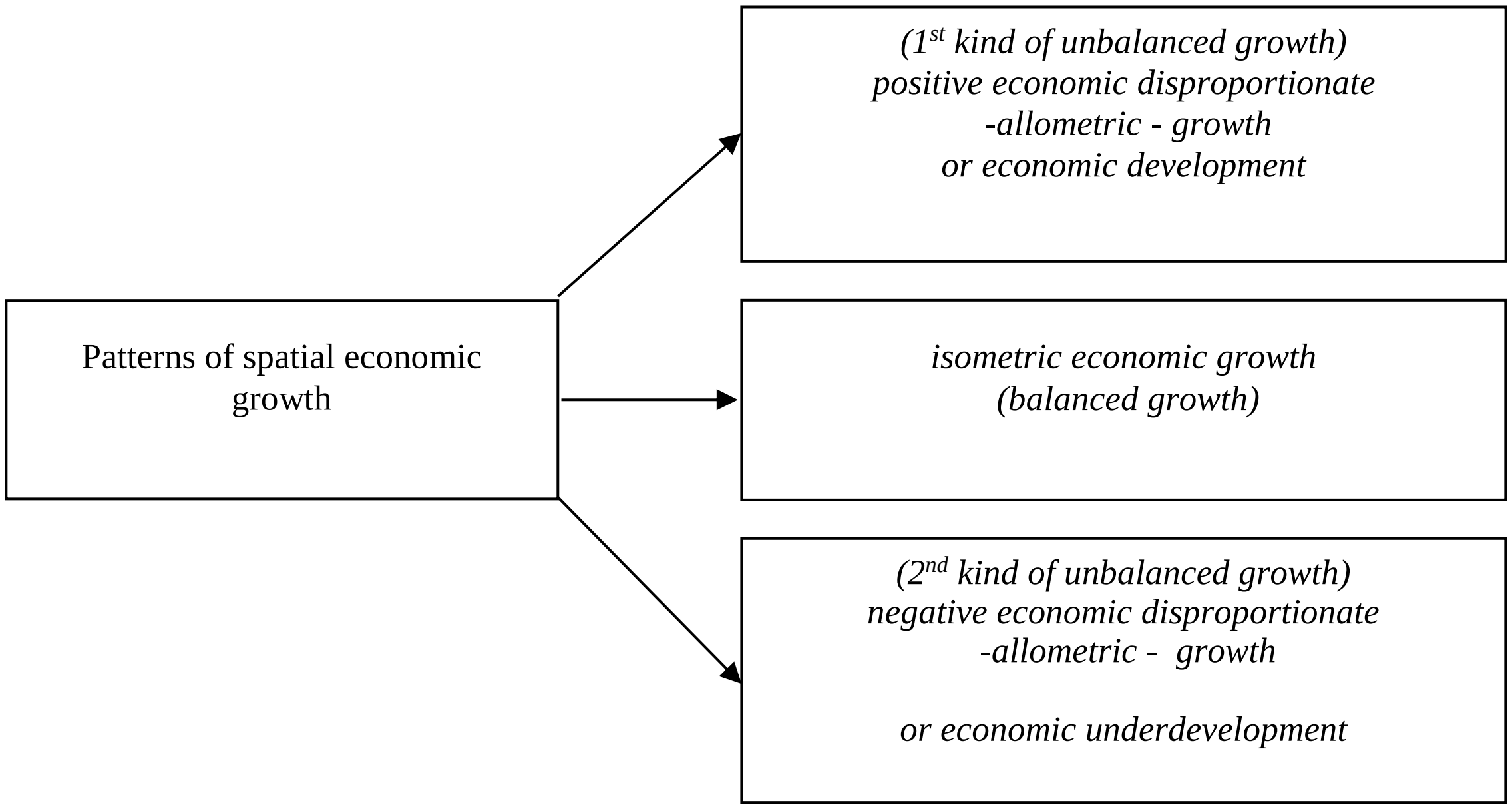

The results are summarized in table 2 above, whereas figure 2 shows the magnitude (speed) of regional economic growth; Figures 3 and 4 reveal the spatial morphology of Italian regional economic growth.

Figure 2 – Magnitude (speed) of the economic growth rate of Italian regions in comparison with the North of Italy, using allometric coefficients

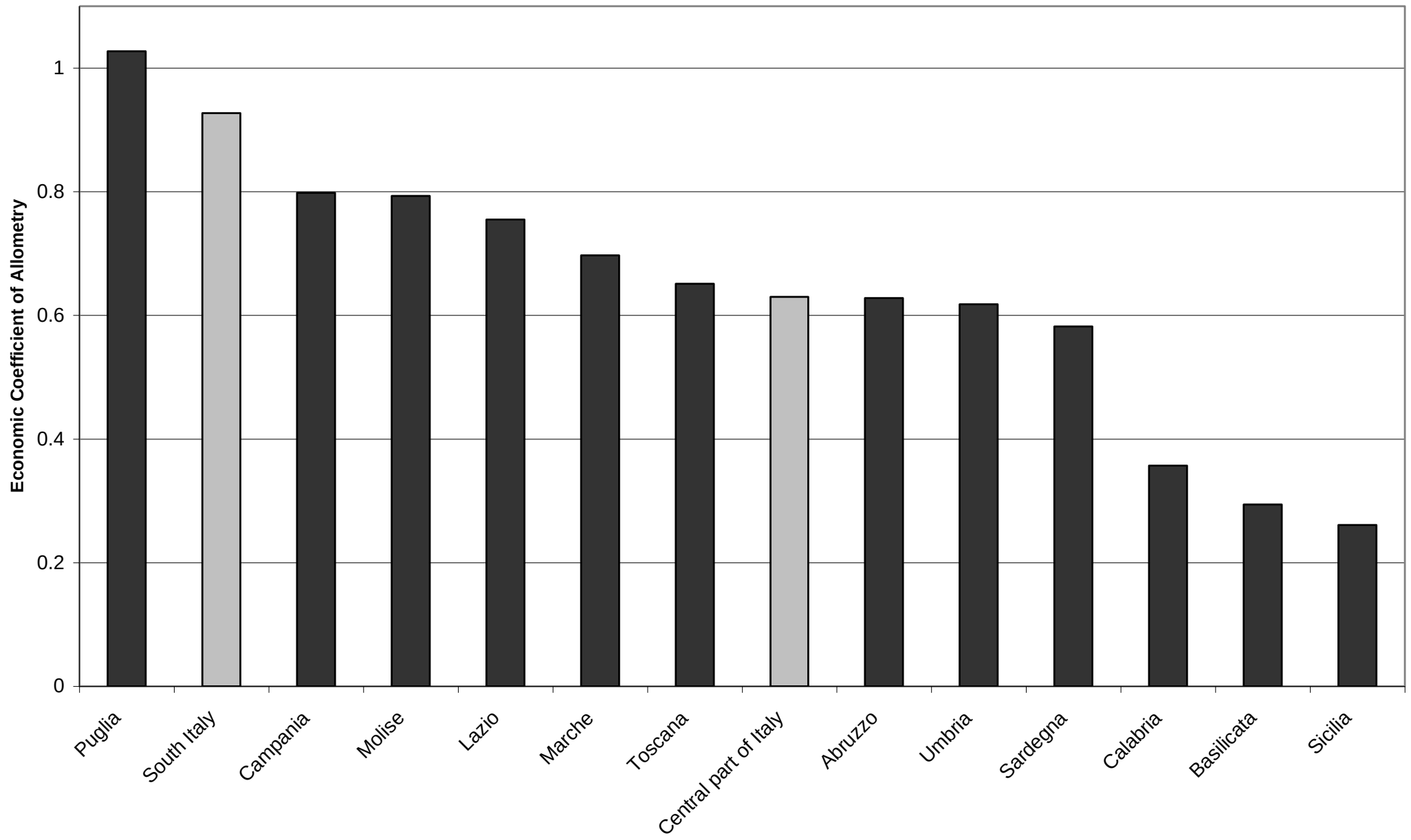

Figure 3 – Patterns of spatial economic growth within Italian regions (1980-2003 period) in comparison with the North of Italy, considering the 5% significance

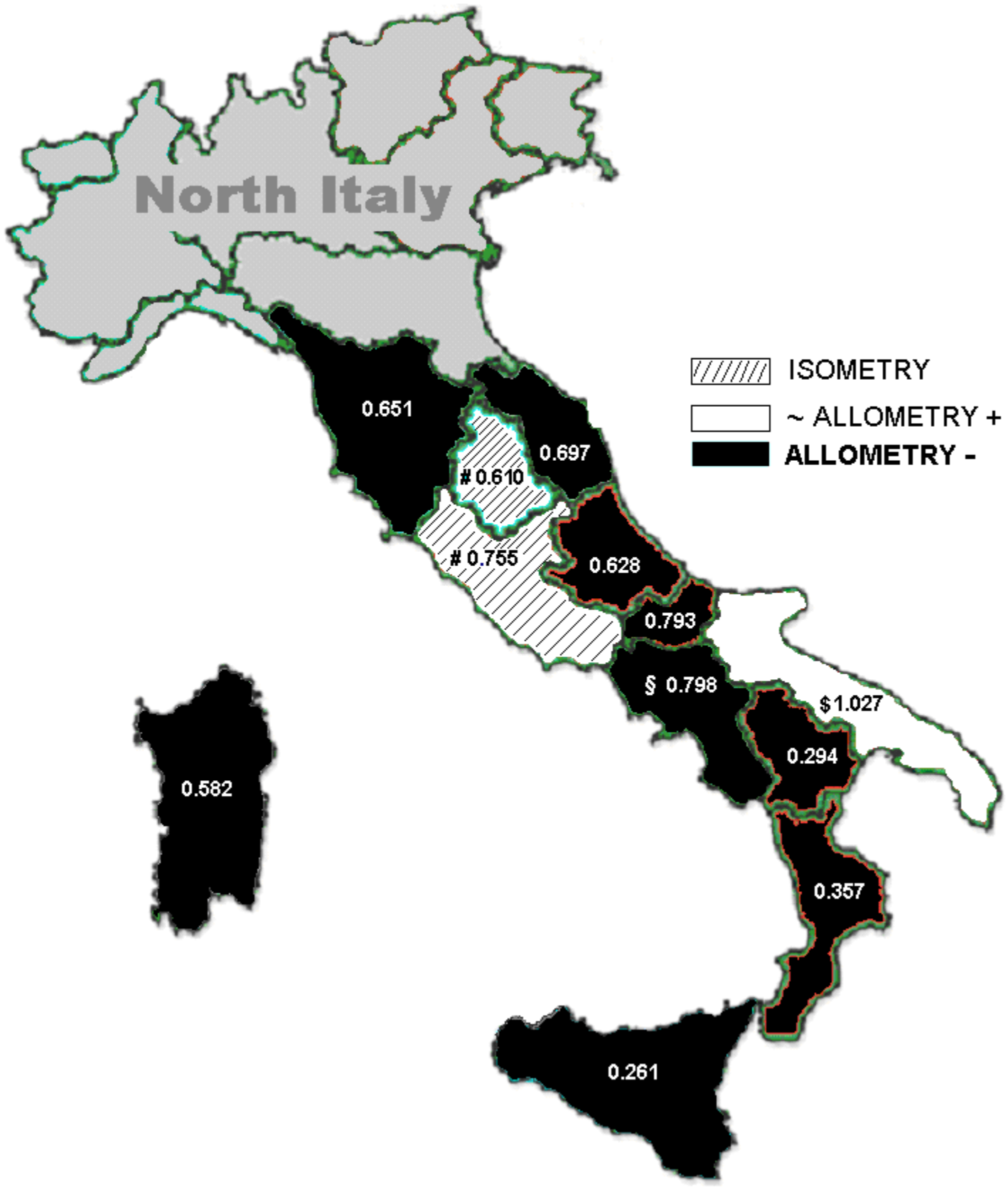

*Note*: Some regions have a non-significant $B_1$, *i.e.* there is no (significant) relationship between the growth rate of the region under investigation and Northern Italy. However, statistical inference on regression coefficients indicates suspect patterns of regional economic growth: **#** Suspect isometric growth in Lazio, Umbria; **§**suspect negative *disproportionate (allometric)* growth in Campania, and **$** suspect positive *disproportionate (allometric)* growth in Puglia.

Figure 4 – Patterns of spatial economic growth of Italian macro-regions (1980-2003 period) in comparison with the North of Italy, considering the 1‰ significance

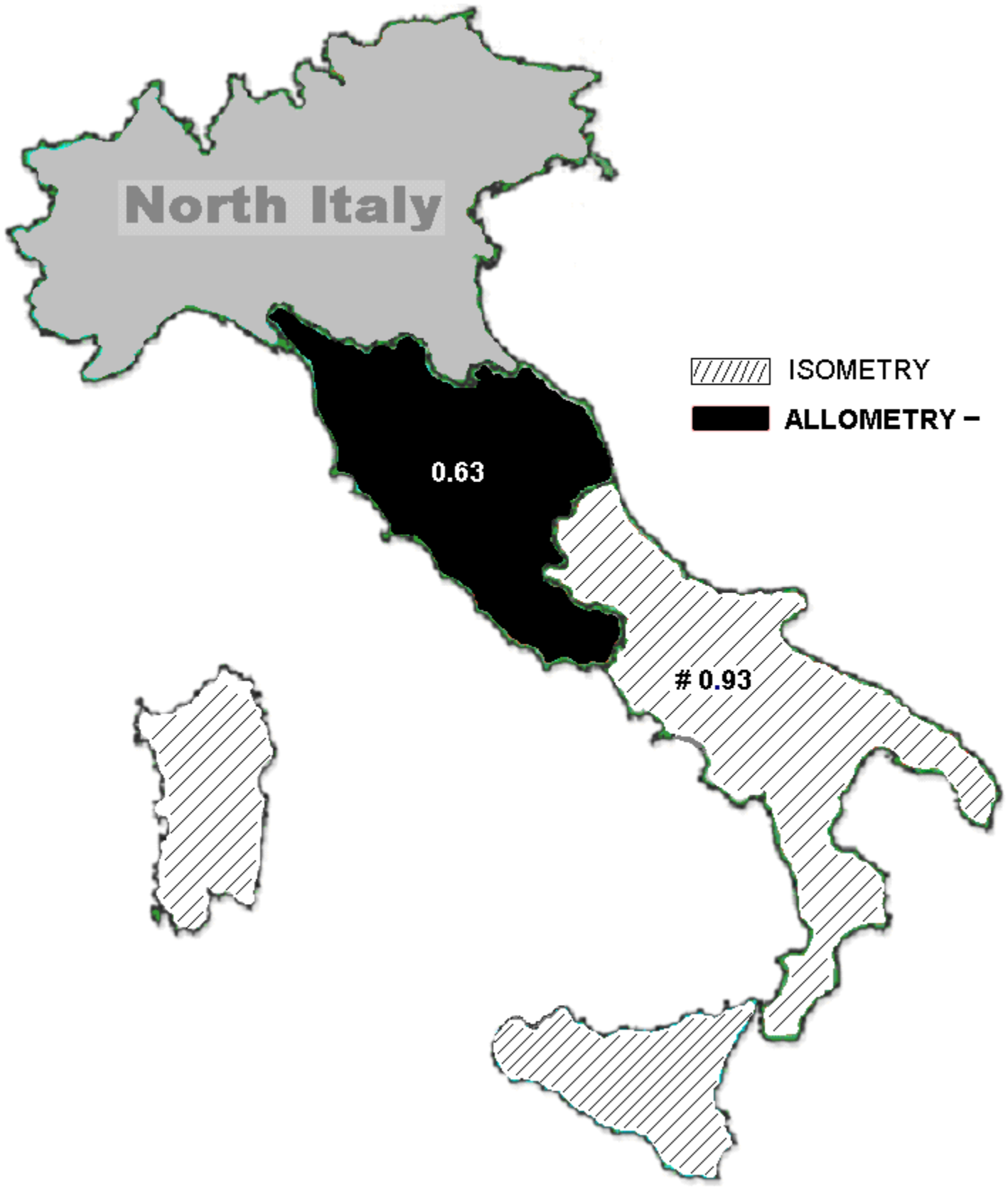

*Note*: The # macro-region has a non-significant $B_1$, i.e. there is no (significant) relationship between the growth rate of the region under investigation and Northern Italy. Statistical inference on regression coefficients indicates suspect isometry in the South of Italy.

Finally, relative changes in economic growth of some regions are represented by the following functions:

$$Pugliay_t = 0.975 \cdot NITALY\, x_t^{1.028}$$

(suspect positive disproportionate – allometric – economic growth)

$$Campaniay_t = 1.185 \cdot NITALY\, x_t^{0.798}$$

(suspect isometric economic growth)

$$Toscanay_t = 0.075 \cdot NITALY\, x_t^{0.651}$$

(negative disproportionate – allometric –economic growth; i.e., under development)

## Appendix

Figure 1A - *S*-shaped economic growth pattern of Central Italy

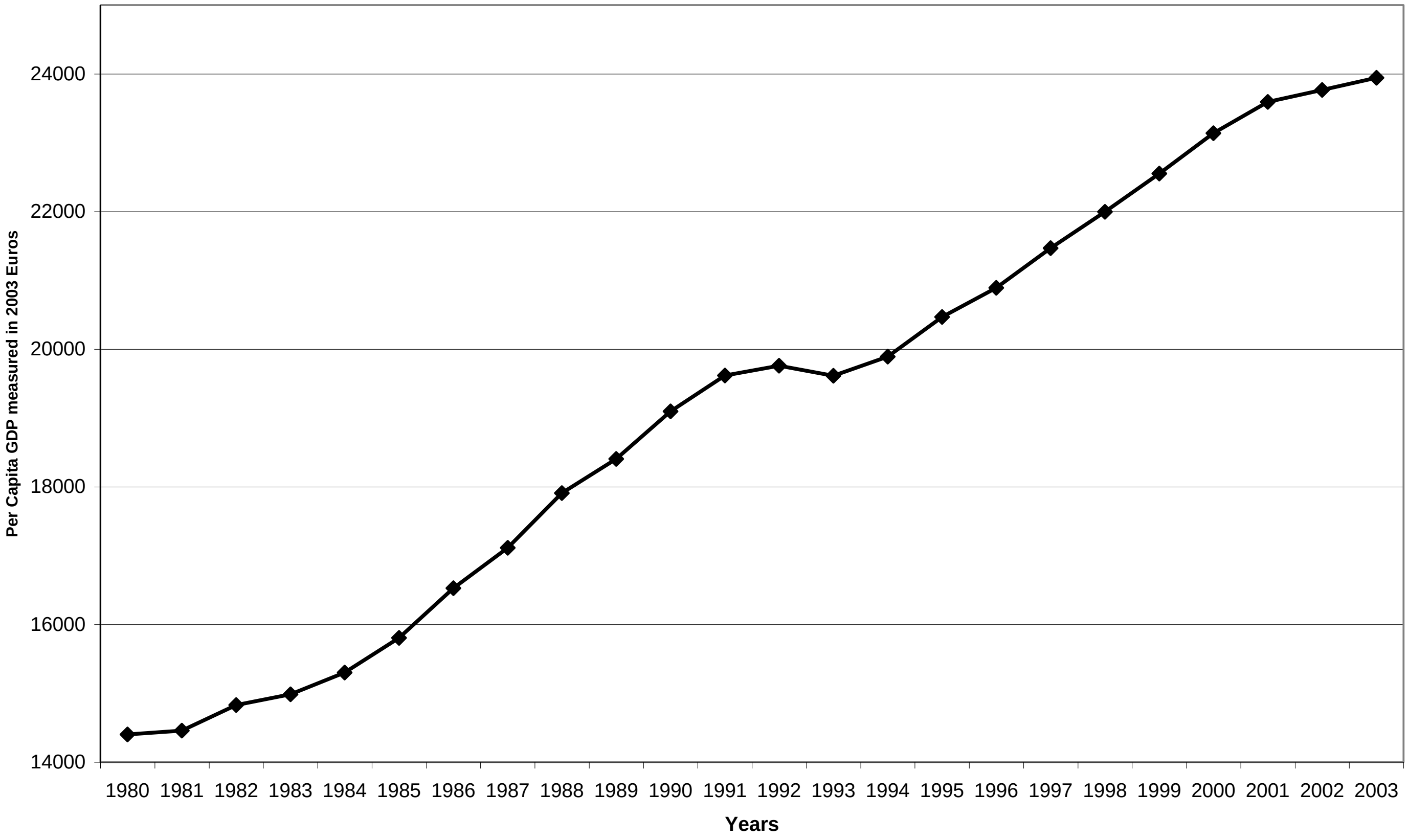

*Note:* This pattern of economic growth is similar in the North Italy, South Italy and other regions

Table 1A - GDP per capita of Italian regions and macro regions in 1981, 1991 and 2001; GDP in millions of Euros, measured in 2003 value

| | | *GDP PER CAPITA* | | |
|---|---|---|---|---|
| | *Regions* | *1981* | *1991* | *2001* |
| North of Italy | Piemonte | 16,439 | 21,645 | 25,871 |
| | Valle d'Aosta | 19,826 | 27,509 | 28,564 |
| | Lombardia | 17,925 | 25,136 | 28,724 |
| | Liguria | 14,070 | 20,272 | 24,818 |
| | Trentino-Alto Adige | 17,861 | 25,942 | 29,120 |
| | Veneto | 15,364 | 21,509 | 25,503 |
| | Friuli-Venezia Giulia | 13,933 | 20,213 | 25,117 |
| | Emilia-Romagna | 18,032 | 23,491 | 28,084 |
| Central part of Italy | Toscana | 15,604 | 20,400 | 24,852 |
| | Umbria | 13,881 | 18,357 | 21,736 |
| | Marche | 13,974 | 18,331 | 22,381 |
| | Lazio | 14,371 | 21,379 | 25,411 |
| | Abruzzo | 11,507 | 17,190 | 19,110 |
| | Molise | 9,895 | 14,337 | 17,372 |
| South of Italy | Campania | 8,958 | 12,536 | 14,873 |
| | Puglia | 9,101 | 12,746 | 14,950 |
| | Basilicata | 8,917 | 11,971 | 15,514 |
| | Calabria | 8,131 | 11,437 | 13,955 |
| | Sicilia | 9,859 | 13,682 | 15,040 |
| | Sardegna | 10,304 | 14,630 | 17,053 |
| | Italy | 13,782 | 19,039 | 22,438 |
| Macro regions | Italy North West | 17,048 | 23,604 | 27,509 |
| | Italy North East | 16,418 | 22,488 | 26,746 |
| | Italy Central part | 14,691 | 20,438 | 24,544 |
| | Italy South | 9,123 | 12,869 | 15,232 |
| | Islands | 9,968 | 13,916 | 15,538 |

Table 2A - Parametric estimates of allometric economic growth model, regions in the 1980-2004 period

| *Regions* | *Estimated relationship* | | | | |
|---|---|---|---|---|---|
| $\ln Abruzzo y_t =$ | 0.406 | $+ 0.628 \ln NItaly\ x_t$ | $R^2$ *adj =0.622* | *F=32.25 (sig.0.00)* | *DW=0.978* |
| *N=20* | (0.131) | (0.110) | *S=0.525* | | |
| $\ln Basilicata y_t =$ | 0.669 | $+ 0.294 \ln NItaly\ x_t$ | $R^2$ *adj =0.036* | *F=1.716 (sig. 0.21)* | *DW=0.994* |
| *N=20* | (0.231) | (0.224) | *S=0.789* | | |
| $\ln Calabria y_t =$ | –2.027 | $+ 0.357 \ln NItaly\ x_t$ | $R^2$ *adj =0.321* | *F=10.927(sig. 0.04)* | *DW=2.207* |
| *N=22* | (0.386) | (0.108) | *S=0.419* | | |
| $\ln Campania y_t =$ | 0.170 | $+0.798 \ln NItaly\ x_t$ | $R^2$ *adj =0.184* | *F=5.518 (sig.0.03)* | *DW=1.436* |
| *N=21* | (0.395) | (0.340) | *S=1.643* | | |
| $\ln Lazio y_t =$ | –2.742 | $+ 0.755 \ln NItaly\ x_t$ | $R^2$ *adj =0.472* | *F=19.773(sig. 0.00)* | *DW=0.867* |
| *N=22* | (0.606) | (0.170) | *S=0.658* | | |
| $\ln Marche y_t =$ | 0.215 | $+ 0.697 \ln NItaly\ x_t$ | $R^2$ *adj =0.684* | *F=46.510 (sig. 0.00)* | *DW=0.392* |
| *N=22* | (0.102) | (0.102) | *S=0.396* | | |
| $\ln Molise y_t =$ | 0.353 | $+ 0.793 \ln NItaly\ x_t$ | $R^2$ *adj =0.855* | *F=130.471 (sig. 0.00)* | *DW=1.331* |
| *N=23* | (0.080) | (0.069) | *S=0.358* | | |
| $\ln Puglia y_t =$ | –0.025 | $+ 1.028 \ln NItaly\ x_t$ | $R^2$ *adj =0.689* | *F=43.124 (sig. 0.00)* | *DW=1.805* |
| *N=20* | (0.156) | (0.156) | *S=0.498* | | |
| $\ln Sardegna y_t =$ | –0.295 | $+ 0.582 \ln NItaly\ x_t$ | $R^2$ *adj =0.213* | *F=6.407 (sig. 0.02)* | *DW=0.932* |
| *N=21* | (0.298) | (0.230) | *S=0.862* | | |
| $\ln Sicilia y_t =$ | 0.735 | $+ 0.261 \ln NItaly\ x_t$ | $R^2$ *adj =0.234* | *F=6.780 (sig. 0.02)* | *DW=1.289* |
| *N=20* | (0.118) | (0.100) | *S=0.482* | | |
| $\ln Toscana y_t =$ | –2.590 | $+ 0.651 \ln NItaly\ x_t$ | $R^2$ *adj =0.744* | *F=62.017(sig. 0.00)* | *DW=1.973* |
| *N=22* | (0.295) | (0.083) | *S=0.320* | | |
| $\ln Umbria y_t =$ | 0.534 | $+ 0.618 \ln NItaly\ x_t$ | $R^2$ *adj =0.553* | *F=25.786 (sig. 0.00)* | *DW=1.019* |
| *N=21* | (0.158) | (0.122) | *S=0.456* | | |

Table 3A - Parametric estimates of the economic model of allometry, macro regions in the 1980-2004 period

| *Macro regions* | *Estimated relationship* | | | | |
|---|---|---|---|---|---|
| $lnCItaly(y_t)=$ | –0.505 | $+ 0.630\ lnNItaly\ x_t$ | $R^2\ adj = 0.857$ | *F=126.990 (sig. 0.00)* | *DW=1.170* |
| *N=23* | (0.200) | (0.056) | *S=0.217* | | |
| $lnSItaly(y_t)=$ | 0.127 | $+0.927\ lnNItaly\ x_t$ | $R^2\ adj = 0.778$ | *F=77.612 (sig. 0.00)* | *DW=1.545* |
| *N=23* | (0.121) | (0.105) | *S=0.542* | | |

*Note 1:* The second column is the estimate of the constant. Given below it, in parentheses, is its standard error. The third column is the estimate of β, with below it, in parentheses, its standard error. The fourth column shows the adjusted $R^2$ of the regression and, below the $R^2$ the standard error of the regression. The fifth column displays the results of the Fisher test, to its right the significance. In the last column is the Durbin-Watson test, which is an indicator of autocorrelation of the time series.
*Note 2.* N=shorter time series in some regions are due to the moving average that reduces the time series by one and also to the impossibility of calculating the logarithm for some values.

FURTHER READINGS FOR THE SOURCES OF ECONOMIC GROWTH BY THE AUTHOR